\documentclass[twocolumn, nofootinbib, showpacs, amsmath, amssymb] {revtex4}

\usepackage{graphicx}

\begin{document}

\title{ Corrections to the running of gauge couplings due to quantum gravity }

\author{ Michael~Maziashvili}
\email{mishamazia@hotmail.com } \affiliation{ Andronikashvili
Institute of Physics, 6 Tamarashvili St., Tbilisi 0177, Georgia \\
Faculty of Physics and Mathematics, Chavchavadze State University,
32 Chavchavadze Ave., Tbilisi 0179, Georgia}

\begin{abstract}

Concerning the gravitational corrections to the running of gauge
couplings two different results were reported. Some authors claim
that gravitational correction at the one\,-\,loop level indicates
an interesting effect of universal gravitational decreasing of
gauge couplings, that is, gravitational correction works
universally in the direction of asymptotic freedom no matter how
the gauge coupling behaves without gravity, while others reject
the presence of gravitational correction at the one\,-\,loop level
at all. Being these calculations done in the framework of an
effective field theory approach to general relativity, we wanted
to draw attention to a recently discovered profound
quantum\,-\,gravitational effect of space\,-\,time dimension
running that inevitably affects the running of gauge couplings.
The running of space\,-\,time dimension indicating gradual
reduction of dimension as one gets into smaller scales acts on the
coupling constants in the direction of asymptotic freedom and
therefore in any case manifests the plausibility of this
quantum\,-\,gravitational effect. Curiously enough, the results
are also in perfect quantitative agreement with those of Robinson
and Wilczek.

\end{abstract}





\maketitle

\subsection*{\textsf{  Introduction}}

Treating the general relativity as an effective field theory, that
is, to consider it as an effective low energy approximation to
some as yet unknown fundamental theory of quantum gravity, offers
a way to get round the familiar renormalization difficulties of
general relativity in the low energy regime \cite{Donoghue}.
Namely, based on the linearized theory of general relativity one
can make reliable predictions about the quantum\,-\,gravitational
corrections in the low energy limit ($E \ll E_P$). Following this
way of reasoning, Robinson and Wilczek considered the correction
to the running of gauge coupling due to graviton exchange in the
vertex diagram \cite{RW}. Using the cut-off regularization, they
found gravitationally corrected $\beta$ function of the form
\begin{equation}\label{RWbeta}
\beta(g,E) = -\frac{b_0g^3}{(4\pi)^2}  - {3 g\over \pi }\left({E
\over E_P}\right)^2~. \end{equation} This result indicates that
gravitational correction always works in the direction of
asymptotic freedom, that is, it always diminishes the coupling.
However, subsequent analysis of this problem has cast serious
doubt on the results of paper \cite{RW}. It was found that this
result was a consequence of an incorrect gauge fixing and that at
the one\,-\,loop level there is in fact no gravitational
correction to the Callan\,-\,Symanzik $\beta$ function
\cite{Pietrykowski}. Further gauge invariant study of the
one\,-\,loop effective action for the gravity\,-\,Maxwell theory
confirmed the absence of gravitational correction \cite{Toms}.
This problem was studied further in a diagrammatic approach as
well, analyzing carefully all the relevant one\,-\,loop diagrams
no gravitational correction was found \cite{EPR}. Recently,
claiming that an appropriate approach to this problem ought both
to preserve gauge invariance and keep quadratical divergences from
the gravitational contributions, the old results were revisited by
using the loop regularization method \cite{TW}. This approach
appears to corroborate the conclusions made in \cite{RW}. While
the discussion of this problem in the framework of an effective
field theory approach to general relativity proceeds, we wanted to
draw attention on a recently discovered profound
quantum\,-\,gravitational effect of space\,-\,time dimension
running \cite{AJLLR} that certainly contributes to the running of
gauge couplings. In view of this profound discovery, we will try
to understand the effect it exerts on the running of gauge
couplings. In implementing of this program we will use a simple
analytic expression of running dimension that comes from a fairly
generic consideration without resorting to any particular approach
to quantum gravity.

\subsection*{\textsf{ QG running/reduction of
space\,-\,time dimension }}

Because of quantum gravity the dimension of space\,-\,time appears
to depend on the size of region, it is somewhat smaller than $4$
and monotonically increases with increasing of size of the region
\cite{AJLLR}. We can account for this effect in a simple and
physically clear way that allows us to write simple analytic
expression for space\,-\,time dimension running. Let us consider a
set $\mathcal{F}$ that is understood to be a subset of four
dimensional Euclidean space $\mathbb{R}^4$, and let $l^4$ be a
smallest box containing this set, $\mathcal{F} \subseteq l^4$. The
mathematical concept of dimension tells us that for estimating the
dimension of $\mathcal{F}$ we have to cover it by $\epsilon^4$
cells and counting the minimal number of such cells,
$N(\epsilon)$, we can determine the dimension, $d \equiv
\dim(\mathcal{F})$ as a limit $d = d(\epsilon \rightarrow 0)$,
where $n^{d(\epsilon)} = N $ and $n = l/\epsilon$. For more
details see \cite{Falconer}. This definition can be written in a
more familiar form as
\[ d =\lim\limits_{\epsilon \rightarrow 0} {\ln N(\epsilon) \over \ln {l \over \epsilon}}~.\]
Certainly, in the case when $\mathcal{F} = l^4$, by taking the
limit $d(\epsilon \rightarrow 0)$ we get the dimension to be $4$.
From the fact that we are talking about the dimension of a set
embedded into the four dimensional space, $\mathcal{F} \subset
\mathbb{R}^4$, it automatically follows that its dimension can not
be greater than $4$, $d \le 4$. We see that the volume of a
fractal $\mathcal{F}$ uniformly filling the box $l^4$ is reduced
\[ V(\mathcal{F}) = \lim\limits_{\epsilon \rightarrow 0} N(\epsilon)\epsilon^4
= \lim\limits_{\epsilon \rightarrow
0}n(\epsilon)^{d(\epsilon)}\epsilon^4 ~,\] in comparison with the
four dimensional value $l^4$. Introducing $\delta N =
n(\epsilon)^4 - N(\epsilon)$, the reduction of dimension
$\varepsilon = 4 - d$ can be written as
\begin{equation}\label{opdim} \varepsilon(\epsilon) \,=\, - {\ln
\left(1 - {\delta N(\epsilon) \over n(\epsilon)^4 } \right)\over
\ln n(\epsilon) } \, \approx \, {1 \over \ln n(\epsilon)}
\,{\delta N(\epsilon) \over n(\epsilon)^4}~. \end{equation} In
quantum gravity the space\,-\,time resolution is set by the Planck
length $\epsilon = l_P$. The local fluctuations $\sim l_P$ add up
over the length scale $l$ to $\delta l = (l_Pl)^{1/2}$
\cite{mazia1}. Respectively, for the region $l^4$ we have the
deviation (fluctuation) of volume of the order $\delta V = \delta
l(l)^4$. Thus in quantum gravity we expect the Poison fluctuation
of volume $l^4$ of the order $ \delta V = (l^2/l_P^2)\, l_P^4 $
\cite{Sorkin}. One naturally finds that this fluctuation of volume
has to account for the reduction of dimension\footnote{This
suggestion has been made in \cite{mazia1}, though the rate of
volume fluctuation was overestimated in this paper, see
\cite{mazia2}. Let us also notice that the necessity of
operational definition of dimension because of quantum mechanical
uncertainties (not quantum\,-\,gravitational !) was first stressed
in \cite{ZS}.}. Respectively, from Eq.(\ref{opdim}) one gets, $n =
l/l_P$, $\delta N = l^2/l_P^2$,

\[ \varepsilon = {1 \over \ln {l\over l_P}}\, \left({l_P \over
l}\right)^2~.\] This equation gives the running of dimension with
respect to the size of region $l$. We can write this expression in
terms of energy, $E_P = l^{-1}_P,\,E = l^{-1}$, as

\begin{equation}\label{rundim} \varepsilon(E) = {1 \over \ln \left({E_P \over E}\right)}\,\left({E
\over E_P}\right)^2~.\end{equation} From the derivation of this
equation it is clear that its validity condition is set by $\delta
V = l^2l_P^2 \ll V = l^4;\, \Rightarrow \,(E/E_P)^2 \ll 1$.

\subsection*{\textsf{Gravitational correction to the $\beta$
function}}

Usually, mass independent renormalization procedure allows one to
write a renormalization group equation in a very simple way
\cite{'tHooft}. Bare coupling in the dimensional regularization
approach has the following general form
\[g_B = E^{\varepsilon} \left[g + \sum\limits_{n =
1}\limits^{\infty} {a_n(g) \over \varepsilon^{n}} \right]~.\]
Differentiating this equation with respect to $E$ and recalling
that $\lambda_B$ does not depend on $E$ one finds \[\varepsilon
\left[g + \sum\limits_{n = 1}\limits^{\infty} {a_n(g) \over
\varepsilon^{n}} \right] + E{dg \over dE} \left[1 + \sum\limits_{n
= 1}\limits^{\infty} {da_n(g) \over dg} \,\varepsilon^{-n}\right]
= 0~.\] Since $\beta$ function, $\beta =E\,(dg /dE)$, is analytic
for $\varepsilon \rightarrow 0$, for small values of $\varepsilon$
one can write \[\beta = A + \varepsilon B~.\] Identifying the
coefficients for various powers of $\varepsilon$ we get \[B + g =
0\,,~~~~~~a_1 + A + B {da_1 \over dg} = 0~.\] That is, we have \[A
= -\left( 1 - g {d \over dg}\right)a_1\,,~~~~~~B = -g~,\] and
correspondingly
\begin{equation}\label{beta}\beta = -\left( 1 - g {d \over dg}\right)a_1
-\varepsilon g~.\end{equation} One might wonder about the
$d\varepsilon/dE$ terms, but by using the Eq.(\ref{rundim}) one
easily verifies that such terms can be safely ignored as far as $E
\ll E_P$. So we have a general expression enabling one to estimate
the correction to the $\beta$ function coming from the dimension
running. Using the Eq.(\ref{rundim})
\[\beta = -\left( 1 - g {d \over dg}\right)a_1
-{g \over  \ln\left ({ E_P \over E}\right)}\,\left({E \over
E_P}\right)^2 ~,\] and then an explicit form of $a_1$ function,
from Eq.(\ref{beta}) we find RG equation for the gauge coupling
 \[E {dg \over dE} = -\frac{b_0g^3}{(4\pi)^2} -{g \over  \ln\left ({ E_P \over E}\right)}\,\left({E \over
  E_P}\right)^2~. \] Integrating this equation one finds

 \begin{equation}\label{gravmodrg}{f(E/ E_P) \over g^2(E) } \,-\,  {f(E_0/E_P) \over g^2(E_0) }
\, = \, {2b_0 \over (4\pi)^2}\,\int\limits_{E_0}\limits^{E}
{f(\xi/ E_P) \over \xi}\,d\xi~,\end{equation} where
\[f(x) = \exp\left( 2 \int\limits_{0}\limits^{x} {\xi\,d\xi \over
\ln(\xi)} \right) = e^{2\,\mbox{li}\left(x^2\right)}~.\] The
logarithmic integral $\mbox{li}(x)$ decays monotonically from
$\mbox{li}(0) = 0$ to $\mbox{li}(1) = -\infty$ in the interval $0
\leq x < 1$. Thus, in the limit $E_P \rightarrow \infty$ one
recovers familiar logarithmic running of the inverse coupling.

If we have several gauge couplings, $g_i$, with corresponding
values of $b_{0i}$, the condition that they unify at a common
value $g(E_U)$ takes the form

\[{1 \over g_i^2(E_0) } \,-\,  {1 \over g_j^2(E_0) }
\, = \, {2 \over (4\pi)^2}\,\,{b_{0j} -b_{0i} \over f(E_0/
E_P)}\int\limits_{E_0}\limits^{E_U} {f(x/ E_P) \over x}\,dx~.\]
Comparing this expression with the unification condition without
the gravitational correction, that is $f = 1$, one finds the
following relation between the uncorrected, $E_*$, and corrected
unification scales
 \[ {1  \over
f(E_0/ E_P)}\int\limits_{E_0}\limits^{E_U} {f(x/ E_P) \over x}\,dx
= \ln{E_* \over E_0}~.\] Here we assumed that the couplings at
$E_0$ are not affected by the gravity, which is certainly good
approximation. As $E_0/ E_P \approx 0$ one can take $f(E_0/ E_P)
\approx 1$ with a good accuracy. Because $f(x)$ is a monotonically
decreasing function one infers that the gravitational correction
increases the unification scale, $E_U > E_*$. In most applications
$E_0 = 10^{-17}E_P\,,\,E_* = 10^{-3}E_P$ \cite{GUT}. To get an
idea how large is the gravitational correction to the $E_*$, one
has to study the solutions of equation

\[\int\limits_{10^{-17}}\limits^{E_U/E_P}
{f(x) \over x}\,dx = 32.2362~.\] We will take less strict but
easier way to get an idea about the gravitational increment of
$E_*$. Estimating the variation of $ E_* = l_*^{-1}$ with respect
to the length fluctuation $\delta l_* = (l_Pl_*)^{1/2}$ one finds

\[\delta E_* = {\delta l_* \over l_*^2} =
{(l_Pl_*)^{1/2} \over l_*^2} = E_*\left({E_* \over
E_P}\right)^{1/2} = 10^{-3/2}E_* ~.\] The value of the
gravitationally corrected coupling at the unification is related
with the uncorrected one, $g_*$, through the relation

\[{f(E_U/ E_P) \over g^2(E_U) } = {1 \over
g_*^2(E_*)}~.\] Finally, let us notice that by omitting a less
important $\ln(E_P/E)$ term in Eq.(\ref{rundim}) we get $f(x) =
\exp(-2x^2)$, so in this case we arrive at the Eq.(\ref{RWbeta})
and one can immediately go along the discussion of \cite{RW}.

\subsection*{\textsf{  Concluding remarks}}

The results derived here are in perfect agreement with those of
\cite{RW}. However, as it was indicated in the introduction, the
paper \cite{RW} is understood to be incorrect for being treated
properly the effective field theory approach to general relativity
does not yield any gravitational correction to the gauge coupling
at the one\,-\,loop level \cite{Pietrykowski, Toms, EPR}.
Nevertheless, we see that quantum gravity through the running of
space\,-\,time dimension \cite{AJLLR} inevitably affects the
running of gauge couplings. Interestingly enough, the results
appear to be in perfect qualitative and quantitative agreement
with those of \cite{RW}. Namely, by taking into account that for
monotonically increasing but otherwise arbitrary expression of
$\varepsilon (E)$ the function $f$ is monotonically decreasing,
one finds that for each value of $E$ there is an intermediate
scale $E_0 < \tilde{E} < E$ such that
\[ {1 \over f(E/ E_P)} \int\limits_{E_0}\limits^{E}
{f(\xi/ E_P) \over \xi}\,d\xi = {f(\tilde{E}/ E_P) \over f(E/
E_P)}\, \ln{E \over E_0}~.\] Denoting $\delta f \equiv
\left.\left[f(E_0/ E_P) - f(\tilde{E}/ E_P)\right]\right/ f(E/
E_P)$ one finds \[{1 \over g^2(E)} = {\delta f \over g^2(E_0)} +
{f(\tilde{E}/ E_P) \over f(E/ E_P)}\,{1 \over g_*^2(E)}~,\] where
we have again assumed $g(E_0) = g_*(E_0)$. Thus on the quite
general grounds, from Eq.(\ref{gravmodrg}) one infers that the
gravitational correction works in the direction of asymptotic
freedom irrespective to the sign of $b_0$. Quantitatively, the
gravitational correction to the Callan\,-\,Symanzik $\beta$
function we have derived above differs from the corresponding
result of \cite{RW}, Eq.(\ref{RWbeta}), by the logarithmic term
$\ln(E_P/E)$ that is not essential in any way. Thus, irrespective
to whether the effective quantum field theory approach to general
relativity indicates or not the gravitational decay of gauge
couplings \cite{RW, Pietrykowski, Toms, EPR, TW}, the profound
quantum\,-\,gravitational effect of spice-time dimension running
\cite{AJLLR} inevitably manifests the presence of this effect. Let
us notice that this sort of investigation inspires interest to
estimate the behavior of gauge couplings in models characterized
with low quantum gravity scale \cite{ADD} as this can be measured
in near future high energy experiments \cite{GL}.

Recently a new interesting paper about the one\,-\,loop effective
action of Einstein\,-\,Maxwell theory with a cosmological constant
appeared in arXive \cite{Toms1}, demonstrating the gravitational
decay of gauge coupling for a positive cosmological constant.
Physically speaking, this result implicitly corroborates our
discussion for the quantum\,-\,gravitational fluctuations of the
background space that results in an effective reduction of
space\,-\,time dimension contributes simultaneously to the dark
energy, that is, to the dynamical cosmological constant
\cite{Sorkin, Sasakura}.

\section*{\textsf{ Acknowledgments}}

The work was supported by the \emph{INTAS Fellowship for Young
Scientists}; \emph{CRDF/GRDF} and the \emph{Georgian President
Fellowship for Young Scientists}.

\end{document}